\documentclass[11pt,twoside]{article}
\usepackage{asp2010}

\resetcounters

\begin{document}

\title{Modeling the Asymmetric Wind of Massive LBV Binary MWC~314}
\author{A. Lobel$^1$,  J. Groh$^2$, K. Torres Dozinel$^3$, N. Gorlova$^4$ , C. Martayan$^5$, 
G. Raskin$^4$, H. Van Winckel$^4$, S. Prins$^4$, W. Pessemier$^4$, C. Waelkens$^4$, 
Y. Fr\'{e}mat$^1$, H. Hensberge$^1$, L. Dummortier$^1$, A. Jorissen$^6$, S. Van Eck$^6$, 
and H. Lehmann$^7$
\affil{$^1$Royal Observatory of Belgium, Ringlaan 3, B-1180 Brussels, Belgium}
\affil{$^2$MPI for Radioastronomy, Auf dem H\"{u}gel 69, D-53121 Bonn, Germany}
\affil{$^3$University of S{\~a}o Jo{\~a}o Del Rei, CAP, 36420-000 Ouro Branco, MG, Brazil}
\affil{$^4$University of Leuven, IvS, Celestijnenlaan 200 D, B-3001 Heverlee, Belgium}
\affil{$^5$ESO, Alonso de Cordova 3107, Vitacura, Santiago, Chile}
\affil{$^6$Universit\'{e} Libre de Bruxelles, Blvd. du Triomphe, B-1050, Brussels, Belgium}
\affil{$^7$Th\"{u}ringer Landessternwarte, Sternwarte 5, D-07778 Tautenburg, Germany}
}

\begin{abstract}
Spectroscopic monitoring with Mercator-HERMES over the past two years reveals
that MWC~314 is a massive binary system composed of an early B-type primary
LBV star and a less-luminous supergiant companion. We determine an orbital
period $\rm P_{\rm orb}$ of 60.85 d from optical S~{\sc ii} and Ne~{\sc i}
absorption lines observed in this single-lined spectroscopic binary. We find an
orbital eccentricity of $e$=0.26, and a large amplitude of the radial velocity
curve of 80.6 $\rm km\,s^{-1}$. The ASAS $V$ light-curve during our spectroscopic
monitoring reveals two brightness minima ($\Delta$$V$$\simeq$$0^{\rm m}$.1)
over the orbital period due to partial eclipses at an orbital inclination angle
of $\sim$70$\deg$. We find a clear correlation between the orbital phases and the detailed shapes
of optical and near-IR P Cygni-type line profiles of He~{\sc i}, Si~{\sc ii},
and double- or triple-peaked stationary cores of prominent Fe~{\sc ii} emission
lines. A preliminary 3-D radiative transfer model computed with {\sc Wind3D}
shows that the periodic P Cygni line profile variability results from an
asymmetric common-envelope wind with enhanced density (or line opacity) in the
vicinity of the LBV primary. The variable orientation of the inner LBV wind
region due to the orbital motion produces variable P Cygni line profiles (with
wind velocities of $\sim$200~$\rm km\,s^{-1}$) between orbital phases $\phi$ =
0.65 to 0.85, while weak inverse P Cygni profiles are observed half an orbital
period later around $\phi$ = 0.15 to 0.35. We do not observe optical or near-IR
He~{\sc ii}, C~{\sc iii}, and Si~{\sc iii} lines, signaling that the LBV's
spectral type is later than B0. Detailed modeling of the asymmetrical wind
properties of massive binary MWC~314 provides important new physical information
about the most luminous hot (binary) stars such as $\eta$~Carinae.
\end{abstract}

\begin{figure}[!ht]
\plottwo{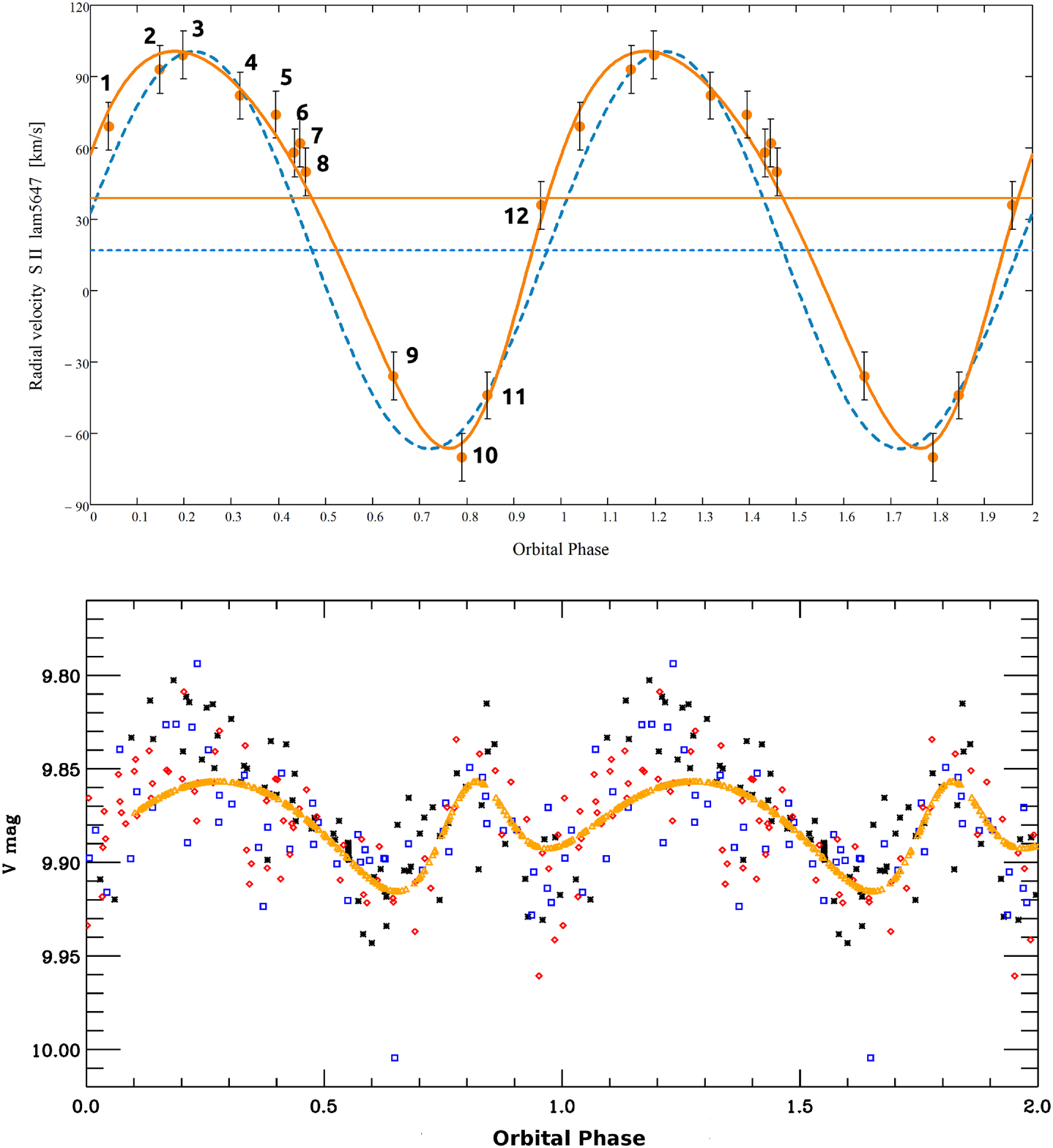}{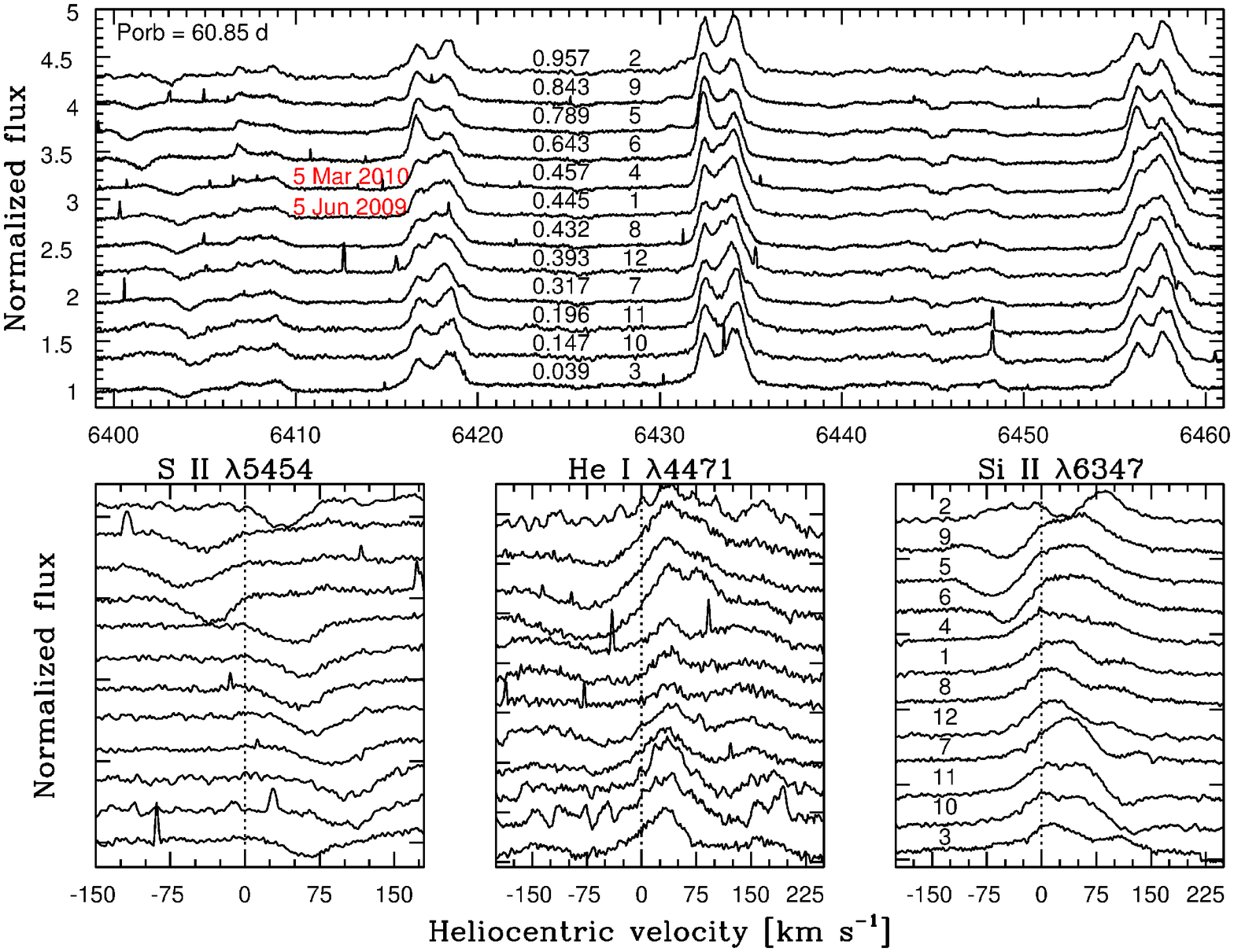}
\caption{$V$- and radial velocity-curve of MWC 314 from high-res. spectra ({\it see text}.)}
\end{figure}

\section{Introduction}
MWC~314 (V1492~Aql; BD$+$14$\deg$3887; $V$=$9^{\rm m}$.9) is a candidate
Luminous Blue Variable (LBV) previously proposed to be one of the most luminous
stars of the Galaxy by \citet{miro1}
with log(${L}_{\star}/{L}_{\odot}$)$\simeq$6.1$\pm$0.3, $T_{\rm eff}$$\simeq$25 to 30 kK, and 
$\dot{M}$$\sim$ $\rm 3\,10^{-5}$ $\rm M_{\odot}\,{yr}^{-1}$. More recently, \citet{mura1}
found that the star is a binary system with an orbital period of $\sim$30
d using optical spectra, however without determining other
orbital parameters. We therefore observed 12 high-resolution spectra over the
past two years with Mercator-HERMES ($R$=80,000) at La Palma (Spain). HERMES is
a high-efficiency {\'e}chelle spectrograph covering 420 nm to 900 nm \citep{rask1}. 
We observed the spectra of MWC~314 with large SNR$\sim$100 for accurate radial
velocity (RV) measurements and detailed line profile studies. On 5 \& 9 Sep 2009, 
and on 17 \& 20 Mar 2011 we also observed two spectra within 5
d to investigate possible short-time spectroscopic variability in MWC~314 
\citep[see also][]{lobe1}.   
  
\section{Radial Velocity and Visual Brightness Curves}
The upper left-hand panel of Fig. 1 shows RV measurements of 
the S~{\sc ii} $\lambda$5454 absorption line in the lower middle panel. 
We find a best fit ({\it orange solid curve}) for an orbital period
$P_{\rm orb}$=60.85 d, $e$=0.26, and systemic (center-of-mass) 
velocity of $+$38.8~$\rm km\,s^{-1}$. The amplitude of the 
RV-curve is large 80.6~$\rm km\,s^{-1}$, and is skewed 
(compare with the blue sine curve) due to the eccentricity of 
the LBV orbit. The lower left-hand panel
shows the $V$-curve observed by ASAS-3 in 2002-2009 \citep{pojm1}. 
The red, blue, \& black dots show 3 epochs of
$\sim$10 $P_{\rm orb}$ folded for one $P_{\rm orb}$. 
The mean $V$-curve ({\it orange dotted curve}) reveals two unequal 
brightness minima due to partial eclipses around periastron 
($\phi$$\simeq$0; the LBV is then in front of the companion) 
and apastron passage ($\phi$$\simeq$0.6). The RV- and 
$V$-curves signal a less-luminous massive (supergiant) companion
star. The shapes of spectral lines in the upper right-hand panel of Fig. 1
very regularly vary over $P_{\rm orb}$. Small flux changes 
in the double-peaked Fe~{\sc ii} emission lines ({\it upper panel}) 
almost exactly repeat over time (compare spectrum Nos. 1 and 4), 
revealing that the optical spectrum variability chiefly results 
from the LBV orbital motion.

\begin{figure}[!ht]
\plotthree{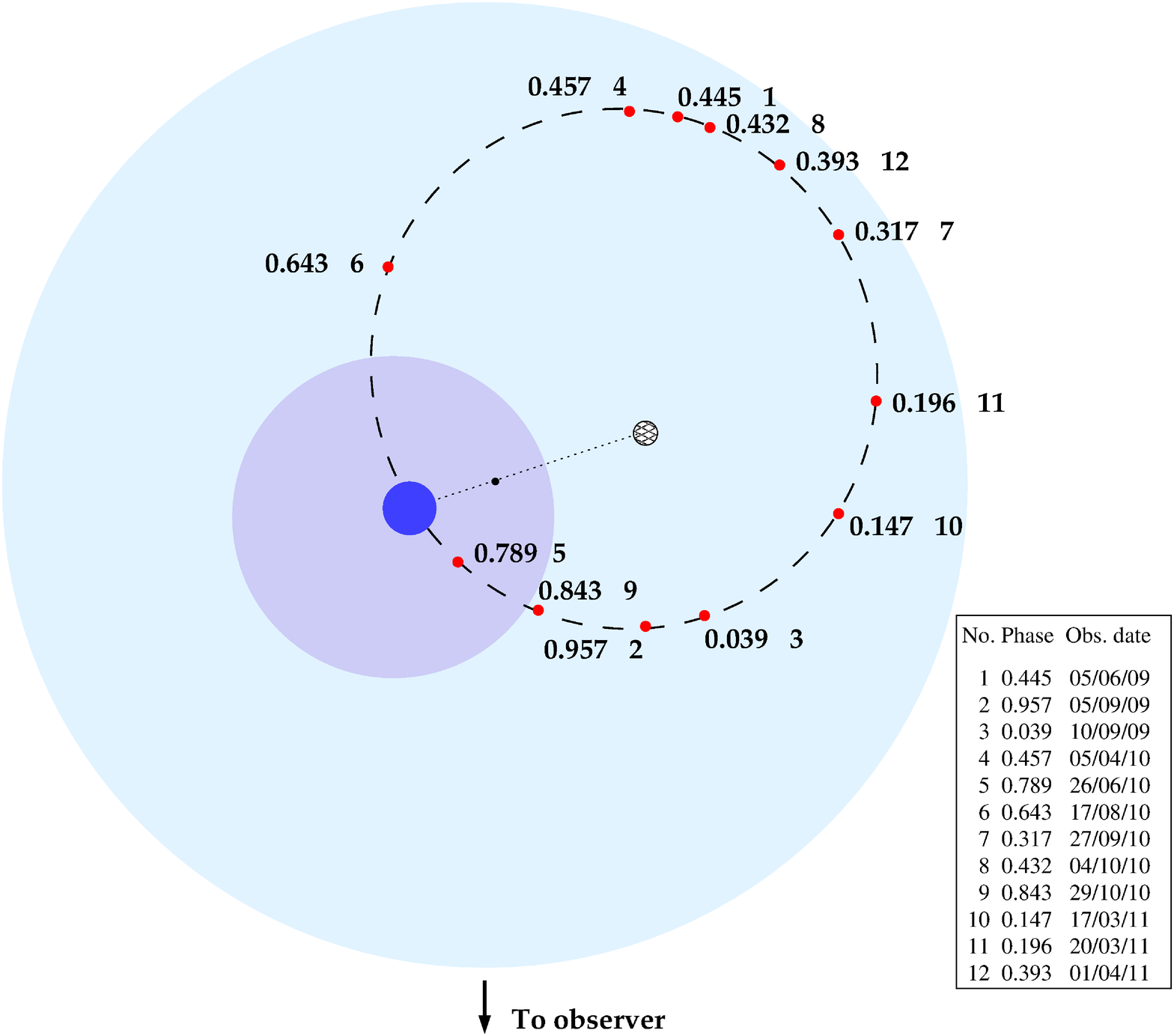}{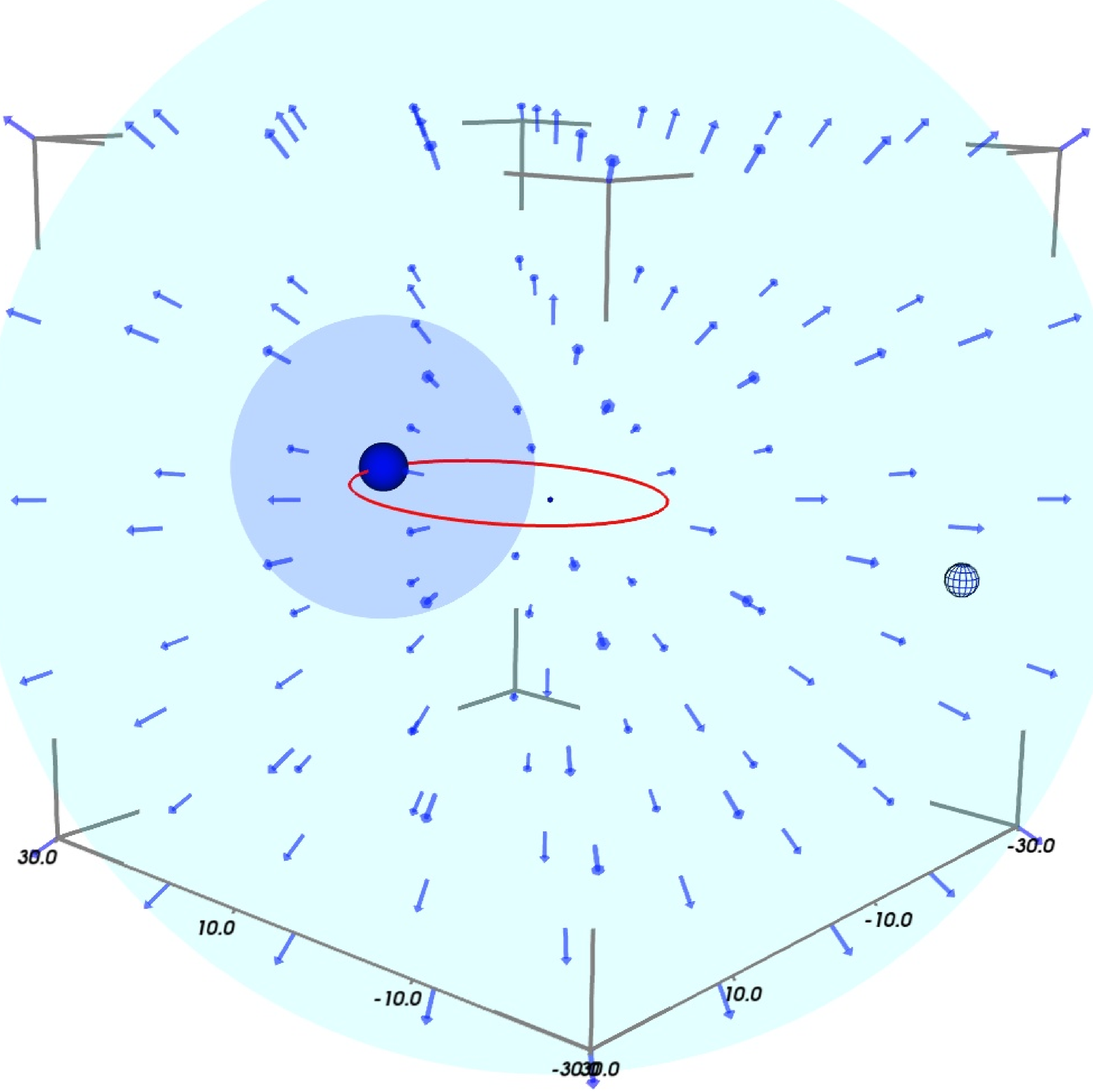}{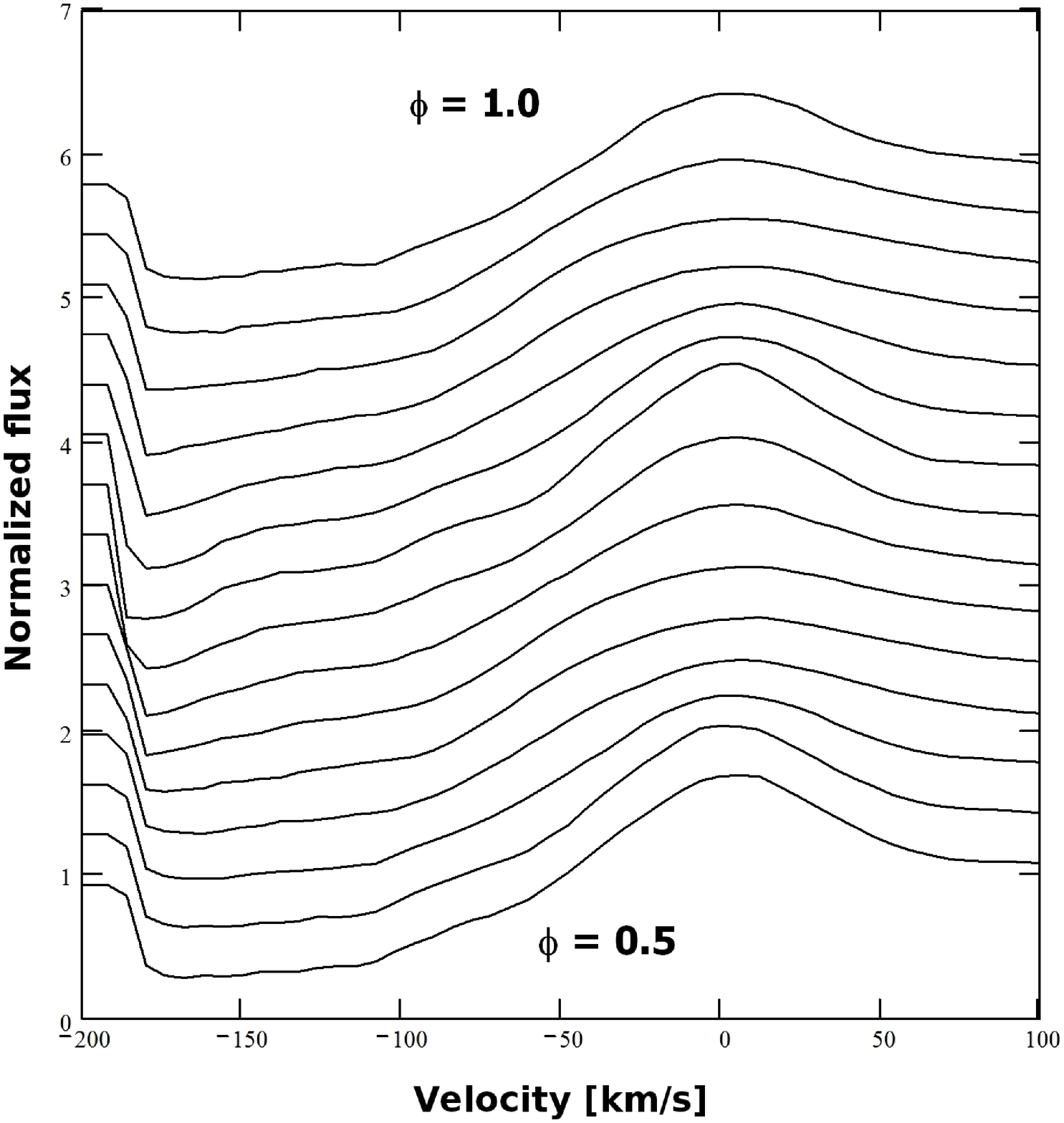}
\caption{Observed orbital phases and 3-D wind model for MWC 314 ({\it see text}).}
\end{figure}

\section{3-D Radiative Transfer Wind Modeling with Wind3D}
The 12 orbital phase positions of the LBV are
shown ({\it solid red dots}) with respect to the companion star in the 
left-hand panel of Fig. 2. Strong P Cyg wind profiles are 
observed for $\phi$=0.65 to 0.85 in He~{\sc i} $\lambda$4471 
and Si~{\sc ii} $\lambda$6347. We compute with the {\sc Wind3D}
code that the P Cyg profiles are properly 
explained with a model of larger wind density surrounding the LBV 
({\it shaded small sphere in the middle panel}) inside a 
circumbinary wind envelope (expanding around the center-of-mass velocity). 
The latter wind envelope causes the static emission lines, 
while the blue-shifted wind absorption becomes most noticeable 
(i.e., He~{\sc I} computed in the right-hand panel of Fig. 2) 
with respect to the line emission when the LBV fastest 
approaches the observer.

\section{Conclusions}
Based on long-term spectroscopic monitoring with Mercator-HERMES we 
confirm the binarity of candidate LBV MWC 314, first conjectured by 
Muratorio et al. in 2008. However, we determine an orbital period of 
60.85 d (twice longer than $P_{\rm orb}$$\sim$1 m they 
proposed) from an accurate solution of the RV-curve. We also compute 
an orbital eccentricity of 0.26 with LBV periastron passage oriented 
almost towards the observer. The visual brightness curve reveals two 
unequal minima signaling partial eclipses at an orbital inclination 
angle of $\sim$70$\deg$ in the plane of the sky. We also confirm 
the LBV character of MWC 314 with strong P Cygni-type line profiles 
observed during the orbital phases of fastest approach around periastron 
passage. A 3-D radiative transfer model we compute for the wind of 
MWC~314 shows that the P Cyg profiles result from enhanced LBV 
wind density inside a circumbinary expanding wind envelope.

\acknowledgements A.L. acknowledges funding from the ESA/Belgian Federal 
Science Policy in the framework of the PRODEX programme. 
The HERMES project and team acknowledge support from the Fund for
Scientific Research of Flanders (FWO), Belgium, support from the Research 
Council of K.U.Leuven (Belgium), support from the Fonds National Recherches 
Scientific, Belgium (FNRS), from the Royal Observatory of Belgium and from 
the Landessternwarte Tautenburg (Germany).


\end{document}